# GENERATION OF GLOBAL VEGETATION PRODUCTS FROM EUMETSAT AVHRR/METOP SATELLITES

*Francisco Javier García-Haro[1], Manuel Campos-Taberner[1], Beatriz Martínez[1], Sergio Sánchez-Ruiz[1], María Amparo Gilabert[1], Gustau Camps-Valls[2], Jordi Muñoz-Marí[2], Valero Laparra[2], Fernando Camacho[3], Jorge Sanchez-Zapero[3] and Beatriz Fuster[3]*

[1]Earth Physics and Thermodynamics Department, Universitat de València, Spain.
[2]Image Processing Laboratory (IPL), Universitat de València, Spain.
[3]Earth Observation Laboratory (EOLAB), Paterna, Spain.

## ABSTRACT

In this work we describe the methodology applied for the retrieval of global LAI, FAPAR and FVC from Advanced Very High Resolution Radiometer (AVHRR) on board the Meteorological–Operational (MetOp) polar orbiting satellites also known as EUMETSAT Polar System (EPS). A novel approach has been developed for the joint retrieval of three parameters (LAI, FVC, and FAPAR) instead of training one model per parameter. The method relies on multi-output Gaussian Processes Regression (GPR) trained over PROSAIL EPS simulations. A sensitivity analysis is performed to assess several sources of uncertainties in retrievals and maximize the positive impact of modeling the noise in training simulations. We describes the main features of the operational processing chain along with the current status of the global EPS vegetation products, including details about its overall quality and preliminary assessment of the products based on intercomparsion with equivalent (MODIS, PROBA-V) satellite vegetation products.

## 1. INTRODUCTION

The main purpose of the LSA-SAF is to develop and implement algorithms that allow an operational use of land surface variables taking full advantage of remotely sensed data from EUMETSAT satellites and sensors to measure land surface variables.

Since the end of 2008, the LSA-SAF generates and disseminates Fractional Vegetation Cover (FVC), Leaf area Index (LAI) and Fraction of Absorbed Photosynthetically Active Radiation (FAPAR) from SEVIRI/MSG data for the whole Meteosat disk at two different time resolutions: daily and 10-day. LAI is defined as half the total area of green elements per unit horizontal ground area [1] accounting for the amount of green vegetation that absorbs or scatters solar radiation. These parameters are used as indicators of the state and evolution of the vegetation cover, and have been used in many agronomic, ecological and meteorological models and applications [2-3].

This paper describes the algorithm currently integrated in the LSA-SAF operational system to produce in near real time and on a 10-day basis global LAI, FAPAR and FVC biophysical products from AVHRR/MetOp data. Unlike the approach adopted to produce SEVIRI/MSG vegetation products, the proposed algorithm relies on the inversion of Radiative Transfer Models (RTMs) for the sake of consistency allowing a joint retrieval of vegetation parameters.

The remainder of this work is organized as follows. Section 2 outlines the main components of the proposed multi-output retrieval chain, including the RTM, the machine learning approaches considered and the assessment of the uncertainty of estimates. Section 3 describes the global EPS products along with useful details about its overall quality and outlines preliminary validation results. Eventually, Section 4 draws final conclusions and outlines future work.

## 2. ALGORITHM DESCRIPTION

The procedure for deriving biophysical parameters relied on an two-step hybrid method: (1) run the RTM in direct mode to build a database of reflectance and associated biophysical parameters representing a broad set of canopy parameterizations, (2) train a non-parametric regression model over the generated simulations using different machine learning approaches.

The main goal of the proposed algorithm is the inversion of the PROSAIL RTM with a family of proposed multi-output kernel-based retrieval methods and neural networks. The best method in terms of stability, accuracy and robustness was then implemented into the operational chain for the joint retrieval LAI, FVC and FAPAR maps globally from corresponding EPS surface reflectance data. A general outline of the methodology is shown Fig. 1.

### 2.1. EUMETSAT AVHRR/MetOp

The EPS is Europe's first polar orbiting operational meteorological satellite. EUMETSAT has the operational responsibility for the MetOp satellites, the first of which (MetOp-A) was successfully launched on October 19, 2006, the second (MetOp-B) in September

17, 2012, whereas the launch of the third (MetOp-C) is foreseen for October 2018.

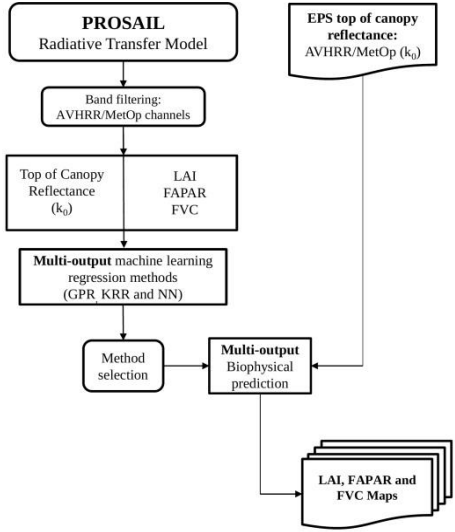

**Fig. 1.** Workflow of the retrieval methodology for the derivation of the EPS LAI, FVC and FAPAR products.

MetOp carries on-board a wide range of sensors, and among them, the AVHRR instrument is the main sensor in charge of providing, providing Earth observations with view zenith angles up to about 60°. This sensor offers capability to observe the whole globe every day at 1.1 km spatial resolution (at nadir). The input of the proposed algorithm is the normalized spectral reflectance factor, i.e. BRDF $k_0$ parameter in three EPS channels, centred at about 0.63 μm (red, C1), 0.87 (NIR, C2) and 1.61 μm (MIR, C3). The algorithm of EPS vegetation products uses as input atmospherically corrected cloud-cleared directional coefficients of the BRDF model, an internal product derived from the albedo algorithm [4].

**2.2. PROSAIL simulations**

The version 5 of PROSPECT and the SAILH were used for PROSAIL RTM coupling. The simulations considered realistic distributions of all leaf and canopy parameters along with bareground spectra representative of all global conditions. A white Gaussian noise was added to the reflectances of the PROSAIL simulations:

$$R_{train}(\lambda) = R_{sim}(\lambda) + \mathcal{N}(0, \sigma^2(\lambda)) \qquad (1)$$

where $R_{train}(\lambda)$ represent reflectance values for band λ used as input in the retrieval algorithm, as obtained adding to PROSAIL simulations, $R_{sim}(\lambda)$, a normal distribution of noise with standard deviation σ(λ).

**2.3. Inversion Methods: multi-output GPR**

In order to invert PROSAIL, we used three powerful non-linear regression methods: the Neural Networks (NNs), and two related kernel-based regression algorithms: the KRR (kernel ridge regression) and the GPR. For the joint retrieval of LAI, FAPAR and FVC we propose multi-output versions.

GPs assume that a multivariate Gaussian prior governs a set of unobserved latent functions, and their likelihood. The observations shape this prior to produce posterior probabilistic estimates. The joint distribution of training and test data is a multidimensional Gaussian and the predicted distribution can be estimated by conditioning on the training data. Standard regression approximates outputs (in or case, the biophysical parameter) as the sum of some unknown latent function f(x) of the inputs (in or case, the normalised reflectance ($k_0$) on the three EPS bands) plus constant Gaussian noise, i.e.

$$y = f(x) + \varepsilon_n, \quad \varepsilon_n \sim \mathcal{N}(0, \sigma_n^2). \qquad (3)$$

A zero mean GP prior is placed on the latent function f(x) and a Gaussian prior is used for each latent noise term $\varepsilon_n$, $f(x) \sim GP(0, k_\theta)$ where $k_\theta$ is a covariance function parametrized by θ and $\sigma_n$ is a hyperparameter that specifies the noise power. Given the priors GP, samples drawn from $f(x)$ at the set of locations $x_i = \{x_i^1, x_i^2, x_i^3\}_{i=1}^N$ (in our case, N is the number of training reflectances simulated with PROSAIL) follow a joint multivariate Gaussian with zero mean and covariance matrix **K** with $[\mathbf{K}]_{ij} = k_\theta(\mathbf{x}_i, \mathbf{x}_j)$. The GP induces a predictive distribution described by the equations:

$$\mu_{GPR*} = \mathbf{k}_*^\top (\mathbf{K} + \sigma_n^2 \mathbf{I})^{-1} \mathbf{y} = \mathbf{k}_*^\top \alpha \qquad (4)$$
$$\sigma_{GPR*}^2 = \sigma^2 + k_{**} - \mathbf{k}_*^\top (\mathbf{K} + \sigma_n^2 \mathbf{I})^{-1} \mathbf{k}_* \qquad (5)$$

where $\mathbf{k}_* = [k(x_*, x_1), \dots, k(x_*, x_N)]$ is an N×1 vector and $k_{**} = k(x_*, x_*)$. The GPR model offers a full posterior probability establishing a relationship between the input and the output variables, from which one can compute pointwise estimations, $\mu_{GPR*}$ and also confidence estimates $\sigma_{GPR*}^2$.

When the goal is to predict multiple variables, the construction of a unique model able to do all the prediction simultaneously may be advantageous, both in computational terms, prediction accuracy and consistency of the predictions.. In order to achieve this goal, the three single output methods considered above have been adapted to derive jointly the three EPS vegetation parameters (i.e, LAI, FAPAR and FVC). This approach was achieved formulating multi-output versions of the NN (NN$_{multi}$), KRR (KRr$_{multi}$), and GPR (GPR$_{multi}$).

In the case of NN, the approach is a multioutput algorithm per se given its characteristics, and relied on the optimisation of the number of hidden layers and neurons, and the learning rate. The optimization of the hyperparameters for the kernel methods was done either by cross validation or by maximizing the marginal likelihood in the case of KRR$_{multi}$ and GPR$_{multi}$. In the standard GPR case we inferred the hyperparameters in $\theta = [\upsilon, \sigma, \sigma_1 ... \sigma_b]$ and model weights using an optimization of the evidence, whereas in the GPR$_{multi}$ we need to optimize the parameters taking into account a global cost function that sumarizes all the cost functions (one per output) into a global cost function.

**2.4 Algorithm Optimization**

We assessed the GPR$_{multi}$ gain in accuracy measuring the reduction in RMSE over test (unseen) PROSAIL EPS simulations with regard to NN$_{multi}$ and KRR$_{multi}$. Results have indicated that important gains in accuracy are obtained for the joint estimation of LAI, FVC and FAPAR, with regard to single-output GPR and the two considered multi-output methods, with a gain in LAI accuracy of 4% and 5% with regard to NN$_{multi}$ and KRR$_{multi}$, respectively, and showing similar improvement in the case of FVC and FAPAR.

In addition, an optimization of the optimal amount of noise indicated that adding moderate noise (e.g. σ=0.015) works reasonably well in all conditions, with overall relative LAI error reductions of 39% for dense green canopies, 30% for dense dark canopies and 25% for intermediate canopies.

**2.5. Products uncertainty estimation**

The algorithm provides an estimate of the confidence assigned at each pixel, taking into account two major sources of error:
  i. The GPR predictive standard deviation ($\sigma_{GPR}$), which quantifies the confidence on the associated estimate.
  ii. A $\sigma_{k_0}$ error, which propagates the effects of the inaccuracies of the BRDF model parameters, on the prediction of biophysical parameters.

### 3. THE EPS GLOBAL PRODUCTS

The multi-output GPR model allows identifying the nonlinear relationship between the three AHVRR/MetOp vegetation products and atmospherically corrected cloud-cleared $k_0$ BRDF product. Fig. 2 shows of the LSA-SAF EPS VEGA (FVC, LAI and FAPAR) 10-day products. The AVHRR based fields are generated pixel-by-pixel, inheriting the temporal and spatial characteristics of the EPS ten-day albedo (ETAL) product. The products are level 3 full globe rectified images in sinusoidal projection, centered at (0°N, 0°W), with a resolution of 1.1km×1.1km. The ETAL is based on AVHRR observations obtained through composite periods of 20 days [4]. We can see that the maps present a good spatial completeness, except in areas usually covered by snow. The is partly due to the recursive temporal composition scheme of $k_0$ inputs maps, which gradually achieve complete spatial coverage after the initialization of the method. All FVC, LAI and FAPAR products are disseminated as a separate file, containing 4 datasets: (1) a vegetation field, (2) an error estimate field, (3) a quality control information field; and (4) the "age" of the information (Z_age) for each image pixel. The products and their respective error estimates are produced using the HDF5 signed 16-bit integer variable. The timeslot in the filename of this product corresponds to the last day of the 20-day time-compositing period.

The algorithm provides an estimate of the confidence or uncertainty assigned at each pixel (see examples of overall error estimates in Fig. 3). Since the first one quantifies how close a pixel is to the training data, inspection of $\sigma_{GPR}$ maps may improve the selection of PROSAIL inputs, such as certain backgrounds not initially included in the preliminary retrieval, and identify possible invalid outliers such as pixels contaminated by traces of snow/ice, undetected clouds or residual atmospheric effects. Beyond certain uncertainty limits (e.g., 0.20 for FVC and FAPAR, and 1.5 for LAI) estimations may be regarded as unreliable and its use should be restricted.

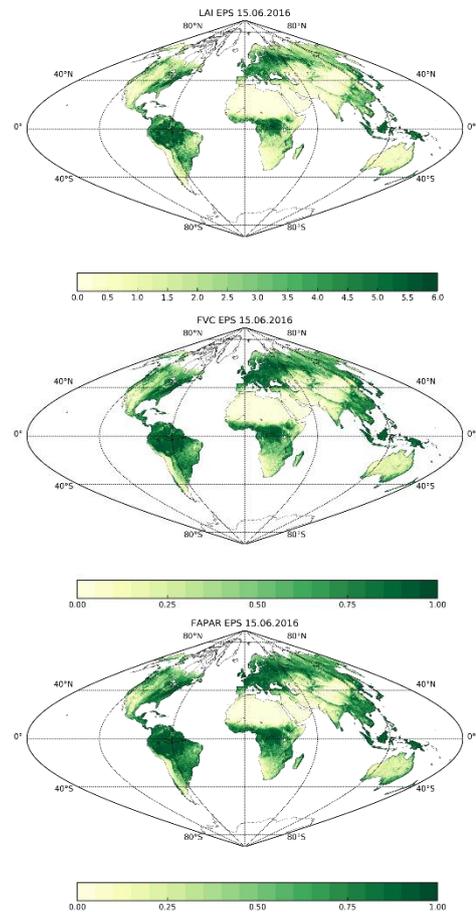

**Fig. 2.** EPS LAI (top), FVC (middle) and FAPAR (bottom) products for 15th June, 2016.

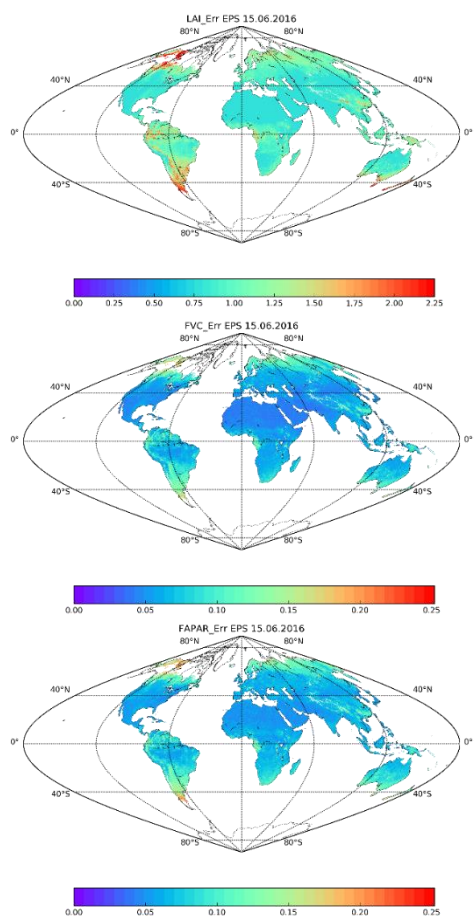

**Fig. 3.** Errors associated to EPS LAI (top), FVC (middle) and FAPAR (bottom) products for 15th June, 2016.

The seasonal variations in the quality and coverage of the products have been also assessed. In overall, around 80% of pixels for all variables showed good (i.e. <0.1 for FVC/FAPAR; <1 for LAI) and medium ([0.10, 0.15] for FVC/FAPAR; [1.0, 1.5] for LAI) quality levels. Only around the 2%, 0.2% and 3.2% of pixels showed poor consistency for FVC, LAI and FAPAR, respectively. Clearly, the best performance for all products corresponds to areas with latitudes below 40ºN, such as in Africa and Australia continents, with optimal quality retrievals in about 90% of the regions and a negligible percentage of poor quality or unprocessed pixels.

The LSA-SAF vegetation products are routinely validated. The adopted strategy for validation of EPS vegetation products consists of three main steps: 1) evaluation of errors in the main variables used as input for EPS algorithm and assessment of the impact on EPS products; 2) inter-comparison with other satellite derived vegetation products; and 3) comparison with in situ measurements. The results of a preliminary assessment of the EPS vegetation products has revealed overall statistical good results compared with references PROBA-V V1 and MODIS (C5 and C6) over a network of sites [5-6].

## 4. CONCLUSIONS

In this work a novel algorithm has been developed for the determination of global vegetation parameters based on data acquired by the AVHRR instrument onboard the satellites of the MetOp series forming the European Polar System. The algorithm was integrated in the LSA-SAF operational system and products are reliably generated and delivered in near real time on a 10-day basis from the LSA-SAF website hosted at IPMA (http://landsaf.meteo.pt).

Estimating several biophysical parameters simultaneously may be beneficial to attain consistent and computationally efficient predictions. The proposed algorithm has demonstrated a good performance and provides also an effective means to reject possible invalid observations. The results of this study have revealed the convenience of adding noise levels to reduce retrieval errors and produce more stable solutions.

Future work will complete the validation studies. The products will be thoroughly assessed to ensure the reliability for all biomes and seasons. The retrieval algorithm will be adapted to take full advantage of the enhanced spectral and directional capabilities of the EPS Second Generation (EPS-SG/VII and EPS-SG/3MI) products. The current validation activities will be continued and extended to Meteosat Thrid Generation (MTG) products with a special focus on inter-calibration and temporal consistency between available families of LSA SAF vegetation products.